%% file: timing.tex
\documentstyle{elsart}

\input epsf

\begin{document}

\begin{frontmatter}

\title{The time structure of Cherenkov images generated 
by TeV $\gamma$-rays and by cosmic rays}

\input{hegra_appt.tex}

\begin{abstract} 
The time profiles of Cherenkov images of cosmic-ray showers
and of $\gamma$-ray showers
are investigated, using data gathered with the HEGRA system of
imaging atmospheric Cherenkov telescopes during the 1997 
outbursts of Mrk 501. Photon arrival times are shown to vary
across the shower images. The dominant feature is a time gradient
along the major axis of the images. The gradient varies with the distance 
between the telescope and the shower core, and is maximal for
large distances. The time profiles
of cosmic-ray showers and of $\gamma$-ray showers differ in a
characteristic fashion. The main features of the time profiles
can be understood in terms of simple geometrical models.
Use of the timing information towards improved shower reconstruction
and cosmic-ray suppression is discussed.
\end{abstract}

\end{frontmatter}

\section{Introduction}

\begin{figure}[htb]
\begin{center}
\mbox{
\epsfxsize12.0cm
\epsffile{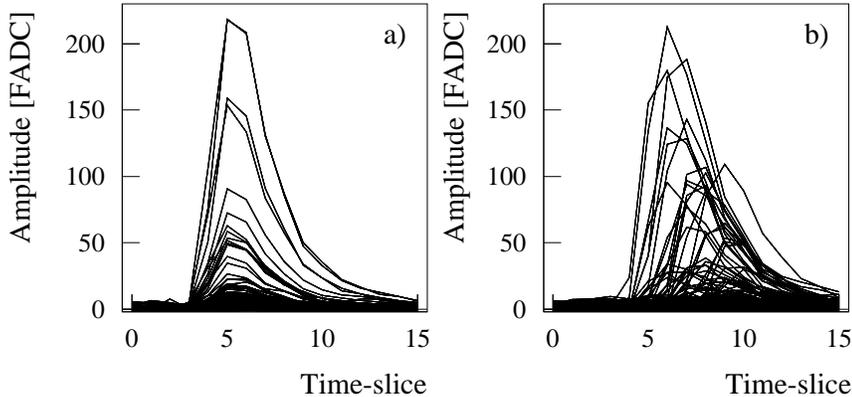}}
\end{center}
\caption
{Signals recorded by the 120 MHz digitizers of the HEGRA IACT cameras,
for two events (a),(b). Signals from all pixels of one camera are superimposed.
One time slice corresponds to 8.3~ns.}
\label{fig_example}
\end{figure}

Imaging Atmospheric Cherenkov Telescopes (IACTs) have become the prime
instrument for $\gamma$-ray astrophysics in the TeV energy range, and a 
number of galactic and extragalactic $\gamma$-ray sources have been 
established using this technique~\cite{overview}. 
Improved understanding of the instruments
as well as methodological advances, such as the stereoscopic 
observation of air showers with multiple Cherenkov telescopes, start to 
allow precision measurements of source properties, as well as increasingly
detailed studies of the relevant characteristics of air showers. 
During the outbursts of TeV $\gamma$-radiation
from the Active Galactic Nucleus (AGN) Mrk 501 in 1997~\cite{501_overview},
the HEGRA 
stereoscopic system of IACTs~\cite{hegra_system,hegra_501}
acquired a large sample of $\gamma$-ray
events, 
ideally suited for such studies
by reason of its purity. The analysis presented here makes
use of one specific feature of the HEGRA telescopes, namely the recording 
of the camera signals using fast digitizers, which sample the 
signal voltages at a frequency of 120 MHz and which allow one
to investigate the pulse shapes and the time structure of Cherenkov images. 
Fig.~\ref{fig_example} shows examples of the signals recorded by the telescopes.
The detailed investigation of the time structure of images 
\cite{hess_phd} was triggered
in particular by the observation of events like the one shown
 in Fig.~\ref{fig_example}(b),
which exhibit an unusually large time dispersion among the signals observed
in different pixels of the IACT camera. Obviously, the timing of pixel 
signals as well as their amplitude
contains information about the evolution of air showers.

In the past, the shape and duration of Cherenkov pulses has mainly been
investigated using angle-integrating detectors, viewing Cherenkov light
from UHE air showers. A seminal review was given
by Patterson and Hillas \cite{hillas1}, and contains further references.
For more recent work, see e.g. \cite{time1} and references given there.
The HEGRA system of Cherenkov telescopes now allows us to study the detailed time
structure of images of TeV cosmic-ray showers and of $\gamma$-ray showers,
resolved into individual image pixels of $0.25^\circ$ diameter, under
conditions where the shower geometry -- direction and core distance -- is
known based on the stereoscopic reconstruction.

The paper is structured as follows: the following chapter introduces 
briefly the HEGRA IACT system and its performance. Then, the
processing of the digitizer signals and the extraction of the amplitude and
timing information is described. After a short overview of the data sample
used in the analysis, results on the time structure of both cosmic-ray induced
showers and $\gamma$-ray induced showers are reported. The paper concludes
with a discussion of applications of the timing information towards improving
the quality of the shower reconstruction, and the rejection of cosmic-ray
showers.

\section{The HEGRA IACT system and its performance}

The HEGRA IACT system is located on the site of
the Observatorio del Roque de los Muchachos
on the Canary Island of La Palma, at a height of 2200 m above sea level.
The IACT system consists of five telescopes, four of them 
(CT2,CT4,CT5,CT6) arranged in the
corners of a square of roughly 100 m side length, with the fifth telescope
(CT3)
in the center. Each telescope has a segmented mirror of 8.5~m$^2$ area
and a focal length of 5~m. During 1997, four of the telescopes 
(CT3,4,5,6) were equipped
with 271-pixel photomultiplier (PMT) 
cameras, with a field of view of $4.3^\circ$
and a  
pixel size of $0.25^\circ$. 
The fifth telescope (CT2) used an older,
coarse-grained camera; this telescope was operated independently of the
four identical telescopes. The same is true for the 
5~m$^2$ prototype
telescope (CT1) near the central system telescope.

Due to the relatively modest reflector size and the large focal
length, the timing properties of the HEGRA mirrors are excellent,
with a maximum time dispersion below 1 ns.

PMT signals of the CT3,4,5,6 telescopes
are sampled and digitized every 8.3 ns
using 120 MHz Flash-ADCs with an 8-bit dynamic range and
a storage depth of 4096 samples. This Flash-ADC memory
holds the signal history of the last 34~$\mu s$. 
To match the bandwidth of the recording system and to suppress frequencies
above the Nyquist frequency of the digitizers, 
the fast photomultiplier signals (about 6 ns fwhm after 22~m of
RG 178 cable)
are shaped to provide a pulse response $u(t) \sim t \exp{-t/\tau}$,
with $\tau = 12$~ns.
After an air-shower
induced trigger, the recording of signals 
with the Flash-ADCs is stopped and a range of
16 samples is read out for each PMT. 
The 16 samples are chosen to cover a range from about 50 ns before the 
arrival of the Cherenkov pulse
to about 80 ns after the pulse.
Only if the digitized
signal exceeds a (software-) threshold of about one photoelectron above 
pedestal, data for a given channel are stored. 
This zero-suppression is required to reduce the data volume; however,
for monitoring purposes, all channels are read out for every 20th event.
The pedestal values for each Flash-ADC channel are updated every
two minutes, based on the few Flash-ADC samples recorded before the
rising edge of a pulse (time slices 0,1,2 in Fig.~\ref{fig_example}).

The trigger signal is generated in a two-stage
process: a {\em camera trigger signal} is generated if the signal in at
least two pixels exceeds a preset threshold of typically 6 photoelectrons. 
The trigger decision is based on the fast, unshaped PMT signals (pulse 
width 6 ns fwhm). To 
suppress random coincidences caused by light of the night-sky background, the
two pixels have to be direct neighbors. Images of $\gamma$-rays showers
are very compact and almost always fulfill this additional condition
\cite{kohler_paper}; the rate
of random coincidences, however, is reduced by almost a factor 50. In a second
stage, two coincident telescope triggers are required to release a {\it system 
trigger}. Adjustable delays compensate the propagation delays which occur 
between
different telescopes, and which depend on the pointing of the telescopes.
Typical rates are in the range of a few 100 Hz to a few kHz 
for individual pixels, 50 Hz to about 500 Hz for single
telescopes, and 15~Hz for the two-telescope coincidence.
To serve as time reference marks, both the local camera
trigger signal and the global system trigger signal are recorded in a 
separate Flash-ADC channel of each telescope.
The trigger system and its performance is described in detail in
\cite{lampeitl_paper}. 

The processing of data involves the extraction of pixel amplitudes as 
described in the following section, and the definition of images, which
include all pixels with a signal 
above a high `tail cut' of 6 photoelectrons, as well
as those pixels above a low tail cut of 3 photoelectrons, which are
adjacent to a pixel above the high tail cut. The image centroid is
defined as the amplitude-weighted
center of gravity of the image pixels, and the image
orientation as the major axis of the image `tensor of inertia'. Also,
the {\em width} and {\em length} parameters are determined as usual
\cite{hillas_param}. 

The image obtained with one telescope corresponds to a view of the air
shower in one projection; by combining the different views obtained 
with the different telescopes, the location and orientation of the
shower axis can be determined by simple geometry \cite{hegra_system}.
The HEGRA telescopes
typically determine the orientation of the shower axis event-by-event 
with a precision of $0.1^\circ$. 
The shower core is located with a resolution of about 10~m,
provided that the core is within about 100~m from the central telescope.
For larger distances, the core resolution worsens, to about 20~m
for 200~m core distance.

The shape of images is used to suppress cosmic-ray showers. Hadronic
showers involve larger transverse momenta than electromagnetic showers
and generate wider, more diffuse images. To provide a single discriminant,
the {\em widths} of all images are divided by the {\em width} 
values expected for 
$\gamma$-ray images,
and an average `scaled width' for all telescopes is formed. 
The expected {\em width} for a given telescope depends on the 
image {\em size} (number of photoelectrons
in the image), and on the reconstructed distance between telescope and
shower core. Other shape parameters are scaled in a similar fashion.
With cuts adjusted to accept about 40\% to 50\% of 
all $\gamma$-rays,
a cosmic-ray rejection factor approaching 100 is possible on the basis of
image shapes alone. Cuts based on the pointing of the
shower axis \cite{hegra_system} provide, for point sources, an additional
(multiplicative) rejection factor of the same order.

In the analysis of shower properties, one has to be aware of a 
potential bias introduced by the criteria chosen to trigger the telescopes
and to select the events used in the analysis. The cuts applied here
to define $\gamma$-ray candidates are quite loose and accept over
80\% of all events, hence a bias is unlikely. The cosmic-ray sample
is more strongly biased; the trigger scheme is designed to enhance
the fraction of `$\gamma$-ray like' events in the recorded data sample,
and some fraction of events is rejected by quality cuts in the
stereoscopic reconstruction. According to our simulations, however,
the measured properties of cosmic-ray showers are quite representative
of the `average' shower.

\section{Processing of digitizer signals}

\begin{figure}[htb]
\begin{center}
\mbox{
\epsfxsize10.0cm
\epsffile{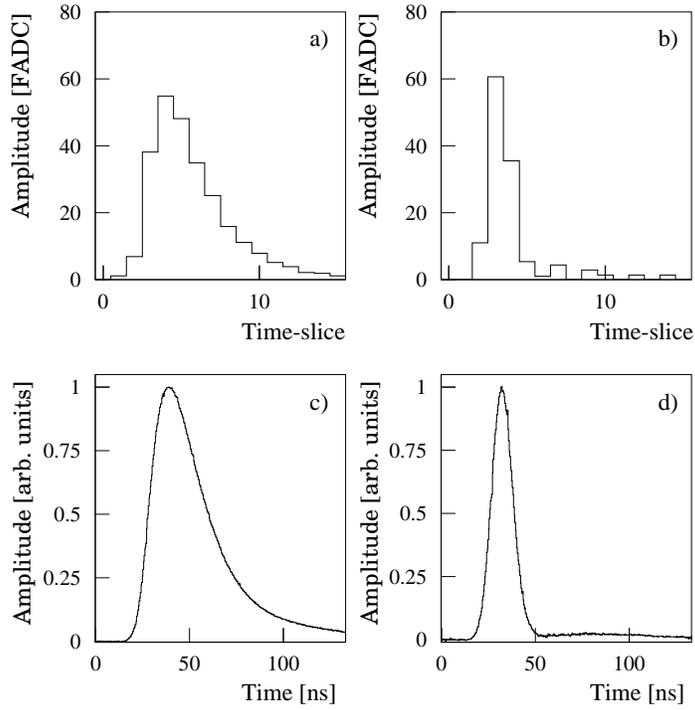}}
\end{center}
\caption
{An individual pixel signal before (a) and after deconvolution (b).
The average signal shape before and after deconvolution is shown in
(c) and (d), respectively, determined by averaging over many signals
and different phases relative to the sampling clock, and plotting the
mean signal as a function of the time difference between the clock 
transition and the signal.}
\label{fig_deconv}
\end{figure}

A first step in the processing of the Flash-ADC signals is the subtraction
of the ADC pedestal; in the next step a digital deconvolution is
applied~\cite{hess_dipl}, which (partly)
reverses the effect of the analog signal shaping. While
this shaping is necessary to allow the reliable recording of signals,
it does increase the effective integration time and therefore the noise
induced by the night-sky background light; the night sky typically contributes
1 photoelectron per 30 ns. In the deconvolution, the new signal $s_k$
in the time slice $k$ is calculated as a linear combination of three
successive
recorded samples $v_k$, $v_{k-1}$, and $v_{k-2}$
$$
s_k = c_1 v_k + c_2 v_{k-1} + c_3 v_{k-2}
$$
with
$$
c_1 = {e^{x-1} \over x},~~~~c_2 = -2 {e^{-1} \over x},
~~~~c_3 = {e^{-x-1} \over x}~~~~.
$$
Here, $x$ stands for $T/\tau$, with the shaping time constant $\tau =12$~ns
and the time interval
between samples $T = 8.3$~ns. Ideally, a delta-function input signal
into the recording chain appears in only two adjacent samples after deconvolution;
the ratio of the amplitudes in the two samples measures the arrival time
of the signal, relative to the Flash-ADC sampling clock. 
In reality one usually finds
small residual entries in adjacent time slices (Fig.~\ref{fig_deconv}(a,b)),
indicating either that the pulse shapes are not ideally matched, or that the 
PMT pulse showed a time spread which is not 
small compared to the shaping time.

A disadvantage of the deconvolution is that the influence of quantization
errors is increased. This is particularly relevant, since
 rather large quantization
steps of about one photoelectron have to be used,
because of the
limited dynamic range of the Flash-ADC system. The benefits of the
significantly reduced integration time (Fig.~\ref{fig_deconv})
are partly offset by the increased quantization noise, 
resulting in a net noise of
1.0 photoelectrons in the pixel signal amplitude, with roughly
equal contributions from the night sky background, and the electronics
noise and quantization.

The arrival time $t$ of the pixel signal is determined based on the center of 
gravity $q$ of the two adjacent peak samples $s_k$, $s_{k+1}$ after deconvolution
$$
q = {s_{k+1} \over s_k + s_{k+1}}~~~.
$$
The quantity $q$ is related to (but not identical to) the phase $\phi$
describing the signal timing relative to the Flash-ADC clock. The
time $t$ is then given by
$$
t = t_k + \phi(q)
$$
where $t_k$ is the time of the clock transition preceeding the signal.
In principle, the relation $\phi(q)$ can be calculated on the basis of
the signal shape and the deconvolution coefficients $c_i$. However,
a more reliable
approach is based on the fact that for air shower events the 
arrival times $t$ are random and hence the distribution in $\phi$ has to
be flat between 0 and 1. Given the measured distribution in $q$, $f(q)$,
one can hence determined $\phi$ using
$$
\phi(q) = \int_0^q f(u) du~~/~~\int_0^1 f(u) du~~~.
$$ 

With this
definition one determines the {\em mean} arrival time of the photons collected
in a given pixel,
for a given event. This time value is not influenced by amplitude-dependent
time-slewing effects, which occur if, e.g., a discriminator is used to
define the pixel timing. Detailed Monte Carlo simulations including the
response of the PMTs, the signal shaping and the digitization by the
Flash-ADC have confirmed that this time determination does reproduce
the average photon arrival time at a pixel \cite{hemberger}.

\begin{figure}[htb]
\vspace{5mm}
\begin{center}
\mbox{
\epsfxsize11.0cm
\epsffile{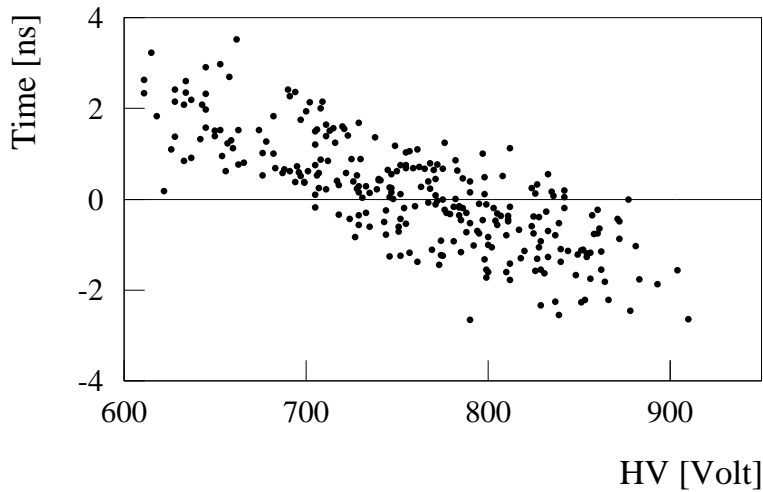}}
\end{center}
\caption
{Mean timing of the light pulser calibration signal, determined for each
pixel, as a function of the high voltage applied to the pixel.}
\label{fig_hv}
\end{figure}

\begin{figure}[htb]
\begin{center}
\mbox{
\epsfxsize11.0cm
\epsffile{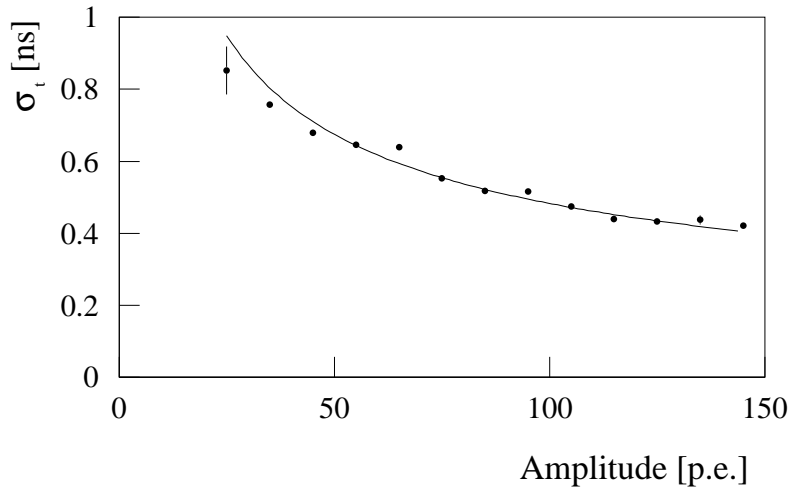}}
\end{center}
\caption
{Time resolution as a function of the signal amplitude in a pixel,
averaged over all pixels.}
\label{fig_timeres}
\end{figure}

Both the relative timing of different pixels as well as the time resolution
was determined by illuminating the camera with a pulsed light source (a
scintillator excited by ns laser pulses) positioned in the center of the
mirror. Small variations in the relative timing of pixels can be caused
by differences in the length of signal cables and of clock cables.
However, the major
effect seen was a dependence of the pixel timing on the high voltage of the PMT 
(the
 voltages are adjusted to provide a uniform sensitivity,
and vary by up to 300 V between PMTs), 
see Fig.~\ref{fig_hv}. The time resolution is
illustrated in Fig.~\ref{fig_timeres}; for signals above 20 photoelectrons,
the time resolution is better than 1 ns. 
The resolution is well described by
$$
\sigma_t = \sqrt{(0.1\mbox{ ns})^2+(4.7\mbox{ ns})^2/A}
$$
with the pixel amplitude $A$ given in units of photoelectrons.

For each event, the arrival times of the signals in the individual
pixels are calculated, and the timing corrections derived using the
light pulser signals are applied. A {\em mean arrival time}
$t_{av}$ for the entire
image is then derived by averaging over those pixels which had a 
signal above the pixel trigger threshold of about 6 photoelectrons.
In this average, pixels which overflow the Flash-ADC dynamic range
are excluded, as are pixels drawing a significant DC current (e.g.,
because a star is imaged into the pixel), and pixels with known faults.

It is not unambiguous how to define the pixel amplitude,
and two options are provided. In one definition, one scans 
the deconvoluted Flash-ADC signal
for the two adjacent time slices with the maximum sum of amplitudes,
within a time window of -8 ns, +20 ns around the
mean arrival time $t_{av}$.
The amplitude $A_t$ is defined as the sum of the two
samples\footnote{Strictly speaking, a
correction needs to be applied, since the reconstructed 
amplitude depends slightly on the signal arrival time
relative to the sampling clock. However,
this correction is quite small, about 4\% rms.}. 
This method has the disadvantage that in particular for 
very small signals the amplitudes are systematically biased towards
larger values, since the selection tends to pick up fluctuations caused
by night-sky background or quantization noise. 

An alternative definition
is to assume that light from the shower will arrive more or less at the
same time in all pixels (at least on the scale of the sampling time,
$T \approx 8$~ns), and to sum for each pixel the two samples closest
to the mean arrival time $t_{av}$, resulting in an amplitude $A_o$. 
This definition results in unbiased amplitudes
for synchronous signals, but may cause a loss of signal amplitude in some
pixels in the case of a genuine timing dispersion among pixel signals
(such as the event shown in Fig.~\ref{fig_example}(b)).
All analyses of HEGRA IACT system data published so far are based on $A_o$.


The effective integration time could in principle
 be further reduced by applying a
`matched filter' to the deconvoluted signals, by weighting the two 
peak samples according to the mean arrival time. For example, 
in case the arrival time
coincides with the clock transition, only a single sample needs to be considered.
However, the gain in noise performance is modest and the sensitivity to 
timing variations is increased further; therefore, this method is 
currently not used.

\begin{figure}[htb]
\begin{center}
\mbox{
\epsfxsize12.0cm
\epsffile{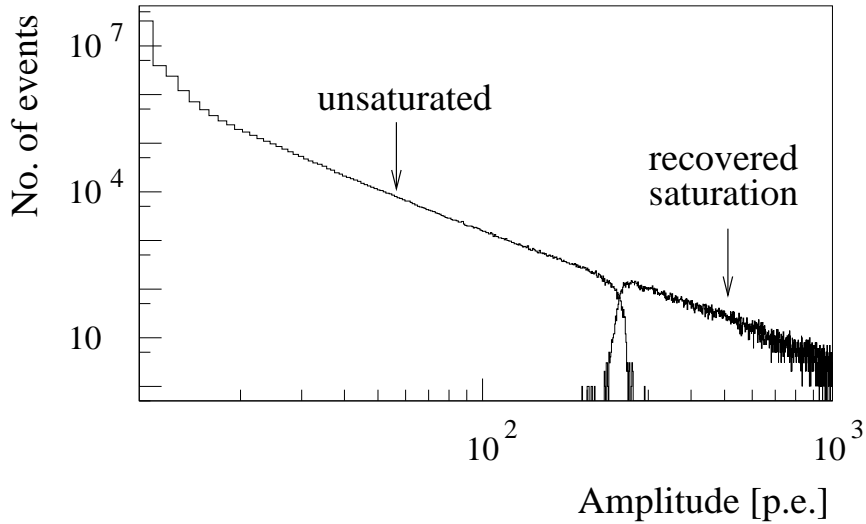}}
\end{center}
\caption
{Spectrum of pixel amplitudes observed for air showers, for unsaturated 
signals (lower part of the spectrum), and for saturated signals after recovery
(upper part).}
\label{fig_over}
\end{figure}

Since the deconvolution procedure can only be applied to linear systems,
pixels which overflow the Flash-ADC dynamic range
have to be handled separately.
Overflow recovery is based on the fact
that for signals well beyond the dynamic range,
the number of samples in the overflow is proportional to the logarithm
of the true amplitude. The length of a saturated pulse, or equivalently
its area, can hence be used to estimate the true amplitude and to
extend the dynamic range. The relation between the area of a saturated
pulse and its true amplitude can be calculated from the pulse shape
determined with unsaturated pulses. Fig.~\ref{fig_over} 
shows the result of this procedure for 
overflow recovery; up to amplitudes of 1000 photoelectrons, the spectrum
of pulse amplitudes looks undistorted and the spectrum at low pixel amplitudes,
where amplitude is directly measured, matches well to the spectrum obtained 
for large amplitudes after the recovery procedure. Beyond 1000 photoelectrons,
nonlinearities in the PMT and in the preamplifier become important, which
cannot be corrected in this manner.

Signal amplitudes were so far always quoted in units of photoelectrons.
The calibration in units of photoelectrons per ADC count is, however,
far from trivial. The method applied to determine the conversion factor
is based on the width of the distribution of pixel amplitudes observed with
the light pulser system. For a pulser signal containing on average $n$ 
photoelectrons (with typical values $n \approx 20$ to 150), the 
width of the amplitude distribution of a given pixel is $\Delta A/A =
1/\sqrt{n}$ in the absence of other effects, 
and one can determine $n$ from this width. In the actual measurements,
one has to correct for effects broadening the signal, such as
fluctuations in the intensity of the light pulses, fluctuations 
$\sigma_{spe}$ in 
the response of the PMT to single photoelectrons, and the (small) 
smearing due to the variable sampling phase. One finds for the relative
width
$$
\sigma^2 = {1 \over n} + {\sigma_{spe}^2 \over n} + \sigma_{phase}^2
+\sigma_{pulse}^2~~~~.
$$
While the first term dominates, there is significant uncertainty in
particular in the determination of $\sigma_{spe}$ from laboratory
measurements and the resulting calibration factor relating ADC counts
to photoelectrons has a systematic error of about 15\%.

\section{Data sample}

The analysis of the time structure of Cherenkov images is based on
an extensive sample of cosmic ray events and $\gamma$-ray events
collected in observations of the AGN Mrk 501 during 1997. A total
of 70 h of data was used. During the observations, Mrk 501 was
positioned $0.5^\circ$ off the optical axis of the telescopes.
Only events at zenith angles between the minimum angle of about 
$10^\circ$ -- given by the location of the source -- and a 
maximum zenith angle of $30^\circ$ were used, resulting in a
mean zenith angle of $17^\circ$. 
To guarantee an optimum reconstruction of
the shower axis, only events with at least three triggered telescopes
were considered. The on-source sample consists of events with shower
directions within $0.2^\circ$ of Mrk 501. To provide a large 
sample of cosmic-ray (background) events, a larger off-source region
was used, consisting of all events reconstructed within $1^\circ$ from the
optical axis of the telescope, but excluding events within $0.4^\circ$
from the direction of Mrk 501. Properties of $\gamma$-ray events are
studied by subtracting, on a statistical basis, on-source and off-source
distributions, the latter scaled according to the solid angle covered 
by the on-source and off-source regions. In the following,
this background-subtracted sample is referred to as the
$\gamma$-ray event sample. After cuts to clean the
event sample, and a cut on image {\em size} between 100 and 400
photoelectrons (see below), about 26400 on-source images
are retained, with an expected background of 8400 cosmic-ray images. 
The off-source cosmic-ray sample includes
177000 images.

As pointed out earlier, a
 potential caveat of the analysis is that the selection
 criterion -- at least three telescopes have to trigger -- 
selects a biased sample of showers. It is well known that such a 
condition, while quite efficient for $\gamma$-ray showers, rejects 
a significant fraction of cosmic-ray showers. Surviving cosmic-ray
showers tend to look `$\gamma$-ray like'. To check for such a possible
bias, timing properties were studied with Monte-Carlo events selected
by different trigger conditions, ranging from all-inclusive samples
to samples with a three-telescope coincidence. Based on these simulations,
one can say that the results concerning time profiles of Cherenkov images are not 
significantly biased by the trigger conditions, and that they represent quite
well the `average' shower.

\section{Time profiles of Cherenkov images}

For each image pixel, the analysis of the raw digitizer data provides
the (deconvoluted) pulse shape, which can be characterized by an 
amplitude, a time (corresponding to the c.o.g. of the pulse), and a pulse
width. The analysis
presented here uses primarily the pixel time; for this quantity,
the experimental resolution of about 1~ns is comparable to the
fluctuations intrinsic to the images. In contrast, the measurement of
the pulse width is limited by the sampling frequency of 120~MHz, 
which does not allow one to resolve the pulse widths of a few ns
characteristic of Cherenkov images.

To characterize the time structure of the Cherenkov images, we consider
the photon arrival time in a pixel
as a function of the location of the pixel
within the image,
averaged over many events. 
Pixel timing, in the following, is always defined as the 
reconstructed time for a given camera pixel, relative to
the mean arrival time $t_{av}$ for the entire camera, as defined above.

\begin{figure}[htb]
\begin{center}
\mbox{
\epsfxsize8.0cm
\epsffile{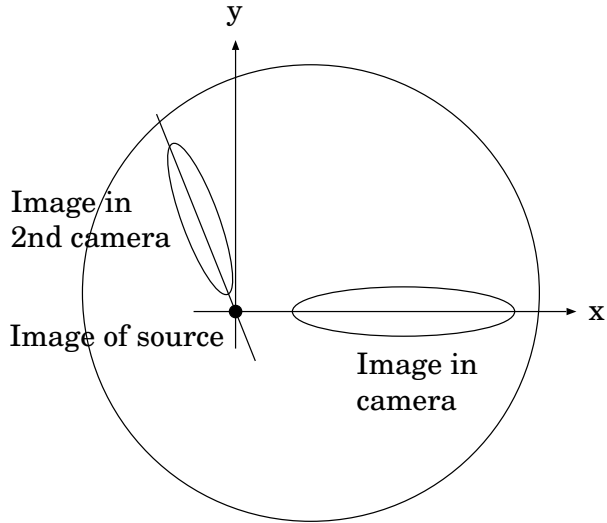}}
\end{center}
\caption
{Definition of coordinate systems.}
\label{fig_coord}
\end{figure}

In order to be able to superimpose images and to average over many events,
images are transformed into a common system, with the origin defined
by the shower direction, i.e., the image of the source, and the $x$-axis
defined by the direction from the image of the source to the centroid of the 
shower image
(Fig.~\ref{fig_coord}) \footnote{Alternatively, one could 
define the $x$-axis as the direction 
from the telescope to the reconstructed location of the shower core in
the telescope plane. For our purposes, the
two definitions give identical results.}. The $y$-axis is then defined to
complete a right-handed coordinate system. The $(x,y)$ coordinates measure
the direction of photons relative to the shower axis, and are
 given in units of degrees. Neglecting the finite
transverse spread of an air shower, the $x$-dependence reflects the
longitudinal evolution of an air shower, since $x \approx r/h$, where
$h$ is the height where a photon is emitted (measured along the shower 
axis), and $r$ is the core distance, i.e., the distance from the telescope
to the shower core in the plane of the telescope dish. Small $x$-values
correspond to emission high up in the atmosphere, large $x$-values to the
tail end of the shower. The $y$-coordinate reflects the transverse spread
of the shower; images are essentially symmetrical in $y$.

The amplitude $A$ and the timing $t$ of a given pixel will depend on the 
location $(x,y)$ of the pixel within the image, i.e. on the direction
of the photons relative to the shower axis, and on the energy and the
impact parameter of an air shower. Here, we are mainly concerned with
the timing characteristics; the variation of image amplitudes with
energy and core distance is discussed elsewhere in detail
\cite{pool}.
The mean photon arrival time at a pixel at a given
$(x,y)$ will primarily vary with the direction under which showers are
viewed, i.e., with the core distance $r$. Variations in shower energy,
or image {\em size} will mainly influence the pixel amplitude, rather
than its timing. Therefore, timing characteristics were studied as
a function of the pixel location $(x,y)$ and of the 
reconstructed core distance $r$,
averaging over a  range of image {\em sizes} between 100 and 400
photoelectrons. It was, however, verified that the results given in the
following do not significantly depend on the choice of the {\em size}
band. 

\begin{figure}[htb]
\begin{center}
\mbox{
\epsfxsize14.0cm
\epsffile{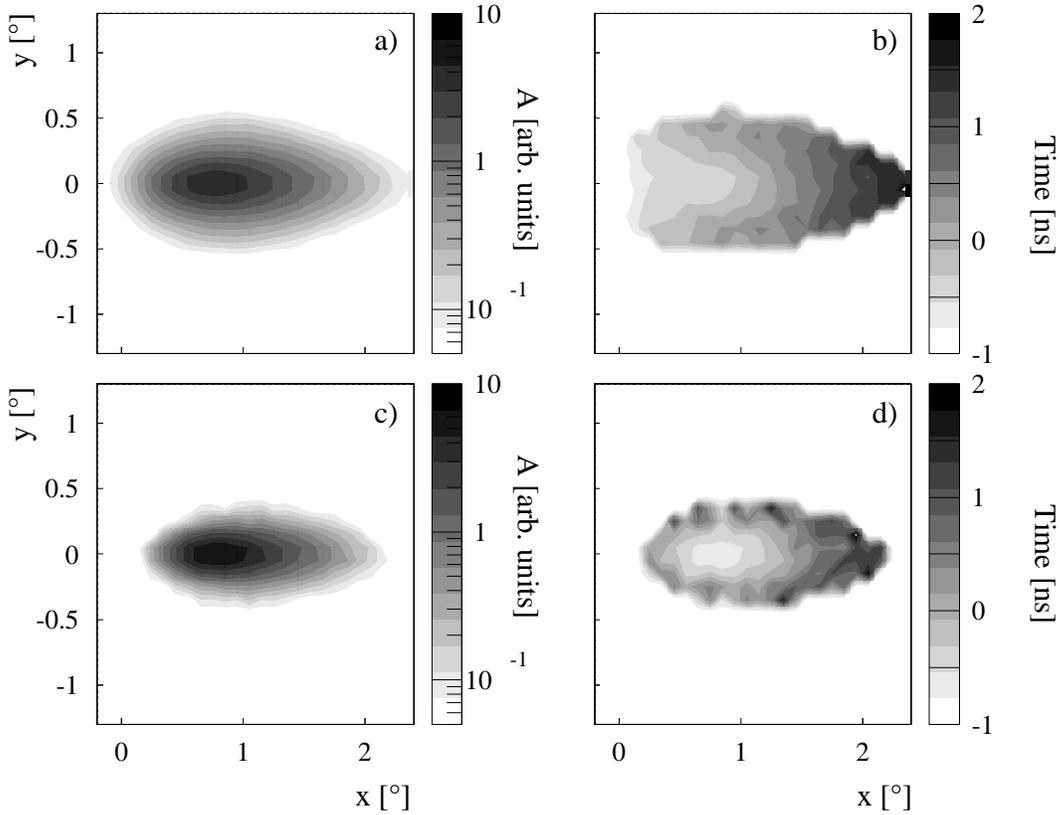}}
\end{center}
\caption
{Amplitude profiles using logarithmic scale (a),(c) 
and time profiles (b),(d) of
hadronic showers (a),(b), and of $\gamma$-ray showers (c),(d), for
reconstructed
core distances $r$ between 100 and 120~m. The profiles were obtained
by averaging over many showers, after transforming the images into
a common coordinate system.}
\label{fig_profile}
\end{figure}

Fig.~\ref{fig_profile} illustrates the amplitude profiles $A(x,y)$
and the time profiles $t(x,y)$ of
Cherenkov images for a reconstructed
core distance between 100 and 120~m, averaged over
many events.

The amplitude profiles $A(x,y)$
show the familiar features; the images are basically
elliptical, but not quite symmetrical in $x$ with respect to the center of gravity
of the images. The $\gamma$-ray images are more compact than the images
of cosmic-ray showers. The dominant feature of the time profiles $t(x,y)$
of hadronic showers
is a gradient along
$x$, with a variation of about 3 ns along the image. The effect seems less
pronounced for $\gamma$-ray showers. Considering the (modest) $y$-dependence
of the time profiles
for fixed $x$, one notes that photons on the image axis 
($|y| = 0$) arrive earliest.

\begin{figure}[htb]
\begin{center}
\mbox{
\epsfxsize11.0cm
\epsffile{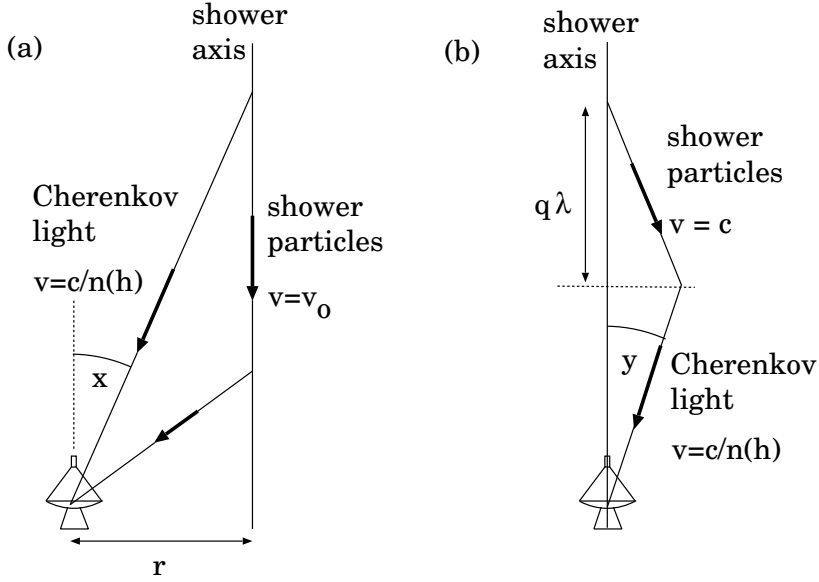}}
\end{center}
\caption
{Illustration of the simple model used to estimate time profiles
(a) longitudinal (along $x$) and (b) transverse (along $y$).}
\label{fig_model}
\end{figure}

Before going into a more detailed 
discussion of the data, it may be worth while to illustrate the key
dependences of image time profiles using an extremely simplified model
(Fig.~\ref{fig_model}).

\begin{figure}[htb]
\begin{center}
\mbox{
\epsfxsize13.0cm
\epsffile{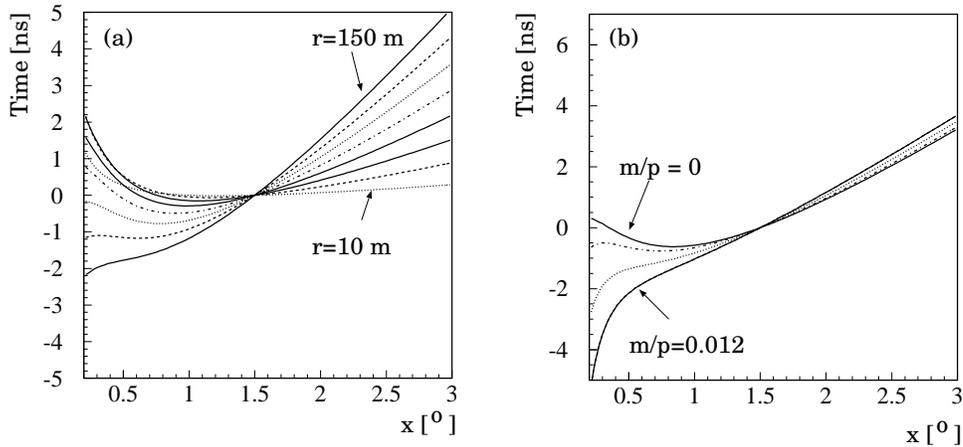}}
\end{center}
\caption
{(a) Longitudinal time profile predicted by the simple model, for 
different core distances $r$ = 10, 30, 50 ... 150 m, 
assuming that the shower propagates at
the speed of light. Positive times imply that the light arrives late.
For each core distance, the origin $t=0$ is arbitrarily defined as
the arrival time for $x = 1.5^\circ$. (b) Influence of the speed of
propagation of the shower front,
$v/c = \sqrt{p^2/(p^2+m^2)}$, for $m/p$ = 0, 0.005, 0.009, and 0.012, 
at a fixed core distance of 100 m.}
\label{fig_model1}
\end{figure}

To describe the dominant longitudinal ($x$) dependence of the timing 
profile, it is assumed that shower particles propagate along the 
shower axis until they reach a height $h = r/x$ where the Cherenkov light 
under consideration is emitted
(Fig.~\ref{fig_model}). The light then propagates to the telescope,
taking into account the variation in refractive index. Fig.~\ref{fig_model1}(a)
shows the resulting longitudinal profiles under the approximation that shower
particles propagate at the speed of light, as justified for electromagnetic
showers. Basically, one finds a gradient across the image, which increases
with increasing core distance $r$. For small $x$, the curves turn over. 
At intermediate core distances,
photons both from very large height and from very small height tend
to arrive late. For large height (small $x$),
light propagates over long distances at a speed $c/n$, which is slightly
less than the speed of the (particle) shower front. These photons lag
 behind the shower front.
For small height, the additional
path length from the shower axis to the telescope causes the delay. The picture
changes once a slower speed of propagation for the (particle) shower front is
assumed, reflecting, e.g., the fact that hadronic showers are 
composed of electromagnetic subshowers, continuously fed by a nucleonic
or mesonic core cascade. For the TeV showers discussed here, characteristic
momenta in this hadronic core of the cascade will be in the range 
of a few 10 GeV to 100 GeV.
For such momenta, the propagation of the shower front is no longer
faster than the speed of light in the medium, and light emitted high up
in the shower (corresponding to small $x$) arrives first (Fig.
\ref{fig_model1}(b)).

As far as the $y$-dependence is concerned, one can obviously no longer
neglect the transverse structure of the air showers. Assuming for 
simplicity (Fig.~\ref{fig_model}(b)) that the shower particles generating
light at a height $h$ and a distance $d \approx y h$ from the shower axis
have their origin on the shower axis at a location $q$ radiation lengths
$\lambda$
above the point of emission, and calculating the $y$-dependence of the
total path length for both the particles and the light, one finds
for the transverse profile
$$
t(y) \approx t_0 + {h \over 2c} \left(1+{h \over q \lambda} \right) y^2
\\
\approx t_0 + \left(2.7 + {18 \over q} \right) y^2 
\mbox{[ns,degr]}~~~.
$$
For the numerical values, 
the typical height of the shower maximum above the HEGRA telescopes,
$h \approx 6000$~m, is used. The transverse time profile
is parabolic, with a variation of typically 1 ns (using $q \approx
2 ... 3$) across the characteristic transverse spread of $0.3^\circ$ of
an image.

\begin{figure}[htb]
\begin{center}
\mbox{
\epsfxsize15.0cm
\epsffile{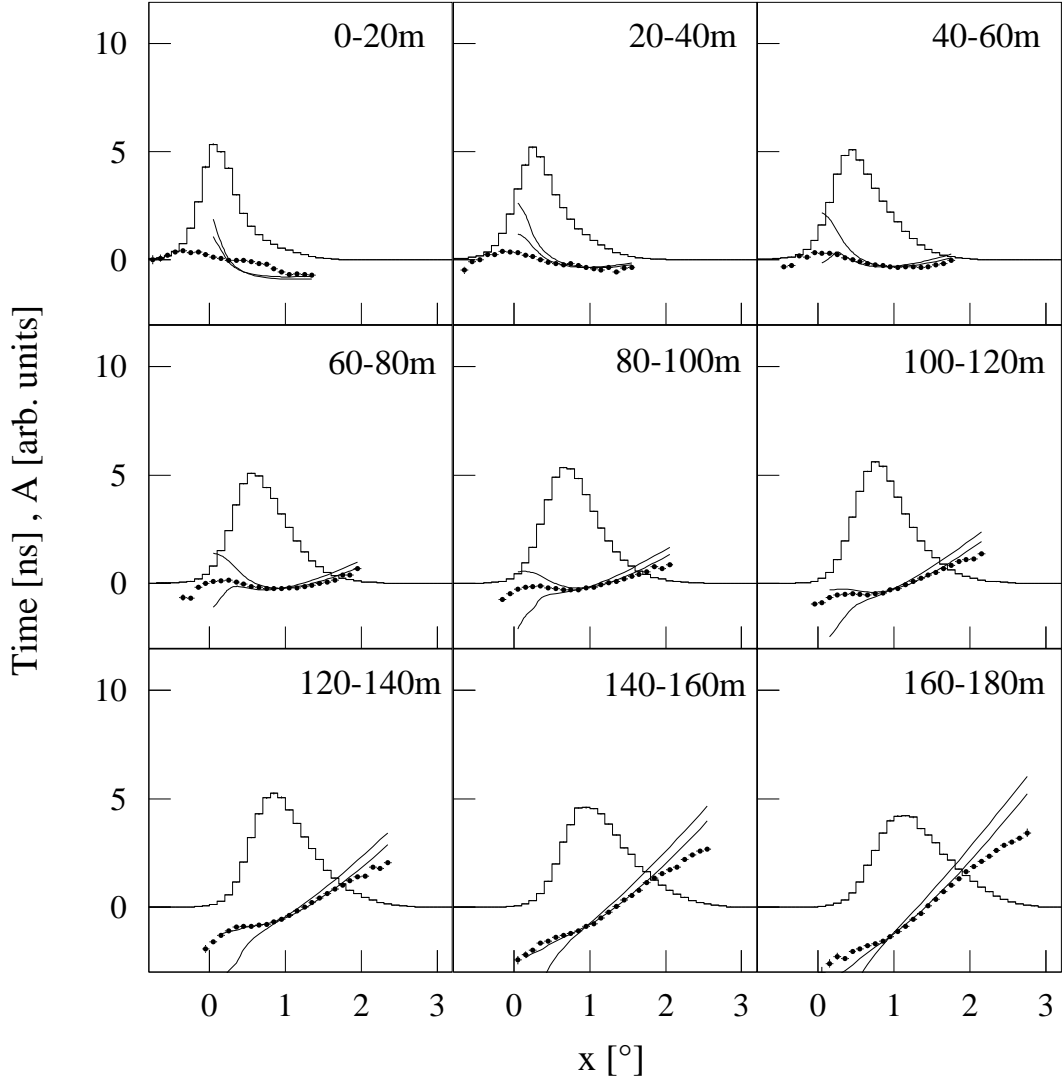}}
\end{center}
\caption
{Longitudinal time profiles $t(x,0)$
of cosmic-ray images, for different ranges
in core distance (filled points). The profiles correspond to a slice
along $x$ through the image, with $|y| < 0.1^\circ$.
Positive times correspond to late arrival of the
photons. Also shown are the corresponding intensity profiles.
The smooth lines refer to the simple model for time profiles, using 
$v/c = \sqrt{p^2/(p^2+m^2)}$, with
$m/p=0.01$ (lower curve, for small x) 
and $m/p=0.0033$ (upper curve) for the shower particles.}
\label{fig_long1}
\end{figure}

Fig.~\ref{fig_long1} shows cuts along $x$ through the time profiles, 
for small $|y| < 0.1^\circ$, for cosmic-ray
showers with different ranges in core distance. These profiles were
obtained by sorting image pixels 
with $|y| < 0.1^\circ$ into bins in $x$ according to the coordinates
of the pixel centers, and
averaging the arrival times in each bin
over large samples of showers. 
For reference, also the
intensity profiles are included, defined via the mean amplitude of pixels
at a given $x$.
In order to avoid a bias in the determination of amplitudes and
arrival times, only those images were used which were well contained within
the camera. The relatively small bins in $x$ used in Fig.~\ref{fig_long1}
should not hide the fact that the measured profiles correspond to the
genuine profile of the Cherenkov light, convoluted with the $0.25^\circ$
pixel size. Structures significantly smaller than the pixel size cannot
be resolved.  Possible systematic effects related to the limited
field of view can be checked by noting that time profiles should only depend on the 
distance $r$ from the telescope to the shower core, but not on the orientation
of the shower axis relative to the telescope axis (since the arrival time of photons
at the telescope location does not depend on the pointing of the telescope).
Indeed, the measured profiles are consistent for shower angles up to
$2^\circ$ from the telescope axis.

The longitudinal amplitude profiles illustrate the well-known
longitudinal asymmetry in the images, the shift of the image 
centroid with increasing distance to the shower core,
and also the increasing {\em length}.

\begin{figure}[htb]
\begin{center}
\mbox{
\epsfxsize10.0cm
\epsffile{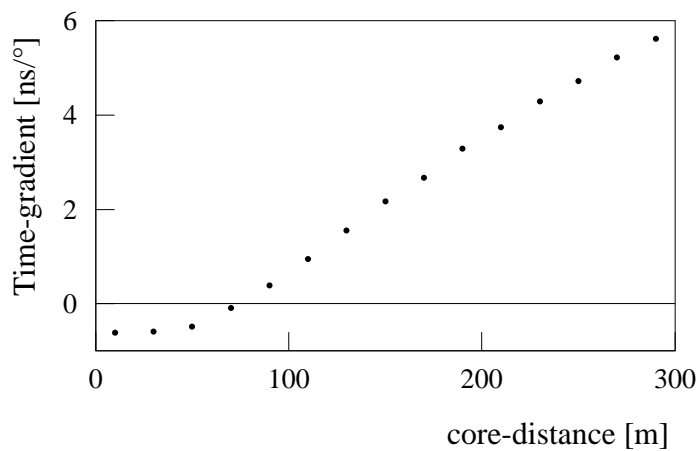}}
\end{center}
\caption
{Longitudinal time gradient of cosmic-ray images, as a function of core
distance. The gradient is determined by fitting a straight line 
to the
data points shown in Fig. 10.}
\label{fig_gradient}
\end{figure}

For these
cosmic-ray showers, one finds -- for a given range in core distances --
an approximately constant
time gradient across the images. The gradient 
varies almost linearly with distance (Fig.~\ref{fig_gradient}), except
for small core distances. For medium core distances, around 100~m to 150~m,
the simple model reproduces  reasonably well the characteristic features of the time
profiles. The deviations observed for small core distances
and small $x$ are not too surprising, since in this domain the transverse
size of the air showers starts to play a role, and $x$ can no longer be 
regarded as a good measure for the height of photon emission.

\begin{figure}[htb]
\begin{center}
\mbox{
\epsfxsize15.0cm
\epsffile{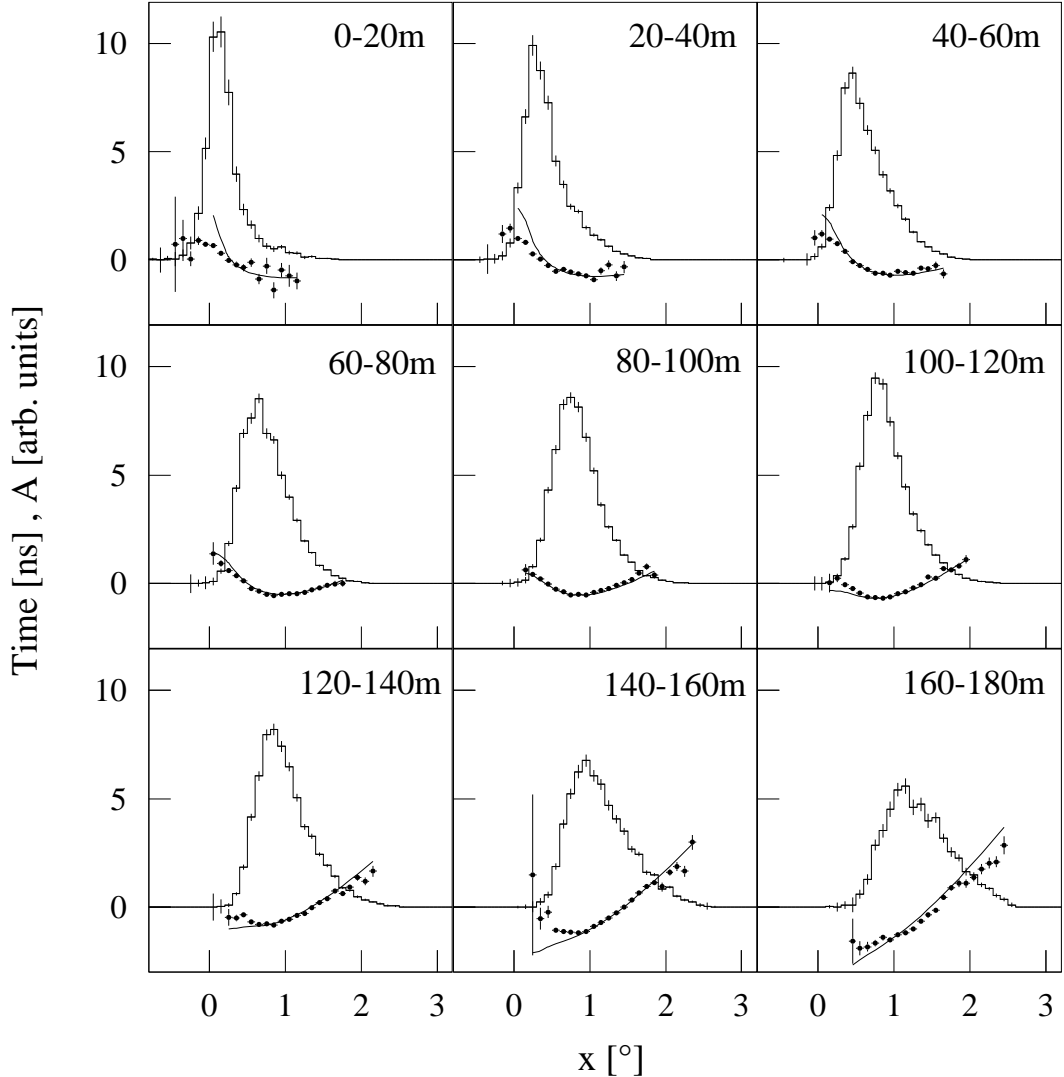}}
\end{center}
\caption
{Longitudinal time profiles $t(x,0)$
of $\gamma$-ray images, for different ranges
in core distance (filled points). 
The profiles correspond to a slice
along $x$ through the image, with $|y| < 0.1^\circ$.
Also shown are the corresponding intensity profiles.
The smooth lines refer to the simple model for time profiles, using 
$v = c$ for the shower particles.}
\label{fig_long2}
\end{figure}

Fig.~\ref{fig_long2} shows the corresponding time profiles for 
$\gamma$-ray showers. Overall, time profiles are similar to the
ones measured for cosmic-ray showers, with a gradient which becomes
increasingly positive for larger core distances. The $\gamma$-ray
profiles differ, however, in some details. In particular, the
profiles in the 60~m to 120~m distance range show a parabolic rather
than linear shape, with photons both from the head and the tail
end of the shower arriving late. This difference between $\gamma$-ray
showers and cosmic-ray showers at small $x$ is expected, and 
reflects the difference in the mass of the shower constituents,
see Fig.~\ref{fig_model1}(b). Except for small core distances, the
simple model with $v=c$ for shower particles does indeed provide a fairly
reasonable description of the measured profiles.

The significant time gradient across the images 
has consequences for the signal acquisition in Cherenkov telescopes. 
In many approaches, a gated charge integrator is used to capture 
pixel signals; to reduce night-sky background, short integration
times of order 10 ns to 20 ns are typically used.
A fixed and very short signal integration time may
result in an underestimate of pixel amplitudes at the head and tail end
of the image, in a lower overall image {\em size}, and a reduced {\em
length} of the image. For the current parameters of the HEGRA telescopes,
with a signal integration time of two time samples or 17~ns after
deconvolution, the differences in image {\em size} are small, 
comparing the
amplitudes $A_o$ for a fixed integration interval, and the
amplitudes $A_t$ with a variable interval. For core distances
below 100~m, the two image {\em size} values differ by less than 2\%;
the difference rises to about 5\% at 200~m and 10\% at 300~m.
For telescopes with larger mirror areas than the HEGRA telescopes,
minimal signal integration times become increasingly important to
reduce the noise of the night-sky background light. From our measurements
it is, however, clear that signal integration times in the 5 to 10 ns regime
 are 
difficult to realize with a common gate applied to all pixels; instead, one needs
either very fast transient recording, or an integration gate which is
adapted to the individual signals.

\begin{figure}[htb]
\begin{center}
\mbox{
\epsfxsize15.0cm
\epsffile{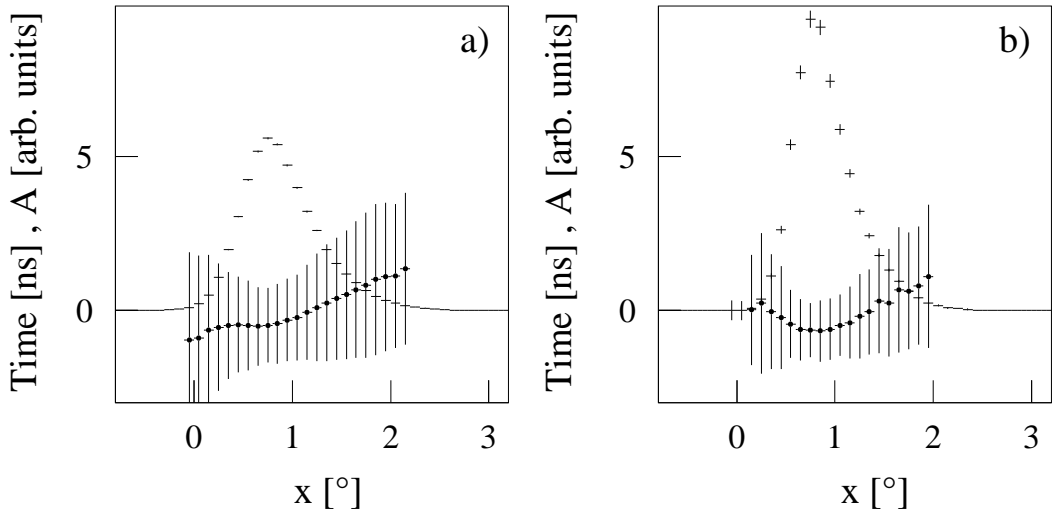}
}
\end{center}
\caption
{Longitudinal time profiles $t(x,\approx 0)$
of cosmic-ray images (a) and of 
$\gamma$-ray images (b) (filled circles), together with the
corresponding amplitude profiles, for a core distance of 100~m
to 120~m. The error bars indicate
the rms width of the distributions of pixel times at each
point along the profile.}
\label{fig_fluct}
\end{figure}

For many applications, not only the time profile of images is of
interest, but also the fluctuations around the average profile, which
are illustrated in Fig.~\ref{fig_fluct}. For the selected class of
images, with {\em size} values between 100 and 400 photoelectrons,
the pixel times fluctuate by 1~ns to 2~ns, depending on the location
of a pixel. The observed fluctuations are about 50\% larger that expected
on the basis of the experimental resolution alone, indicating that
experimental resolution and shower fluctuations contribute about 
equally. Fluctuations seem to be slightly smaller for $\gamma$-ray
showers as compared to cosmic-ray showers. One notes that for
the typical distance range of 100~m to 120~m shown in the figure, the
scale of fluctuations is comparable to the time gradient across the image,
and to the differences between the two shower types.

\begin{figure}[htb]
\begin{center}
\mbox{
\epsfxsize13.0cm
\epsffile{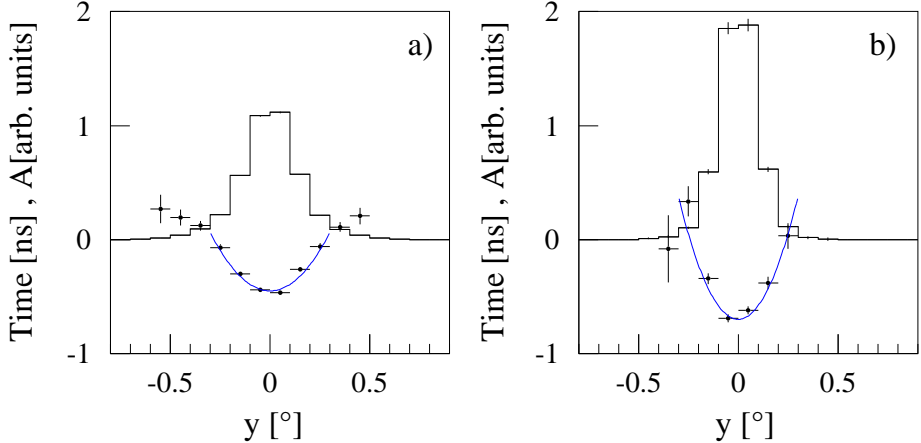}}
\end{center}
\caption
{Transverse time profiles 
$t(x_{max},y)$ of cosmic-ray images (a) and 
$\gamma$-ray images (b), for core 
distances between 100 and 120 m (filled points). 
Also shown are the corresponding intensity profiles.
The smooth lines refer to the simple model for time profiles, which predicts
a parabolic profile.}
\label{fig_trans}
\end{figure}

Fig.~\ref{fig_trans} displays transverse profiles, again
combining data on timing and amplitudes, both for cosmic-ray showers
and $\gamma$-rays. The profiles represent a cut along $y$
at the $x$-value of the peak intensity in the image. 
The width in particular of the amplitude profiles is
of course strongly influenced by the pixel size of $0.25^\circ$.
Close to the image axis, the time profiles show the expected
parabolic shape, $t = t_0 + \beta y^2$, with coefficients 
$\beta = 6$ ns/degr$^{2}$ for
cosmic-ray showers and $\beta = 12$ ns/degr$^{2}$ for $\gamma$-ray 
showers, consistent
with the model prediction of the order of 10 ns/degr$^{2}$.

\section{Applications, including $\gamma$-hadron separation}

In the HEGRA IACT system, the use of Flash-ADCs for signal recording 
was primarily driven by the need to store the signal history for the
about 2~$\mu$s required to generate a trigger based on an inter-telescope
coincidence, and to propagate the trigger signal back to the telescope;
the availability of information on the pixel timing and pulse shapes 
was only a secondary consideration. Nevertheless, having all this
information available, the obvious question is what additional use one
can derive from it. 

Applications can generally be divided into three classes:
\begin{itemize}
\item Improvements in the determination of geometrical shower parameters
such as the shower direction, the core location, or possibly the
height of the shower maximum.
To aid in the determination of shower geometry, one can use
the mean arrival time of each image in the different telescopes, and 
determine a shower direction from the differences in arrival times.
Alternatively, one could use the time gradient across the profile to
estimate the core distance.
\item Improvements in the rejection of cosmic-ray showers.
An additional rejection of cosmic-ray showers might by achieved either
by exploiting the differences in the time profiles between cosmic-ray
showers and $\gamma$-rays, or on the basis of differences in the 
fluctuations of photon arrival times.
\item Consistency checks concerning the interpretation of shower images.
Rather than trying to improve the reconstruction or identification
of shower events, the timing information could be used to provide
on a statistical basis an 
independent confirmation, e.g. of the $\gamma$-ray nature of a given
sample of showers.

\end{itemize}

Here, two selected issues will be addressed, namely the use of the
time gradient over the image for the shower
reconstruction, and the hadron rejection based on the time structure
of images. The use of the average event timing for the reconstruction
of the shower direction was discussed in \cite{daum_timefit}.

\begin{figure}[htb]
\begin{center}
\mbox{
\epsfxsize11.0cm
\epsffile{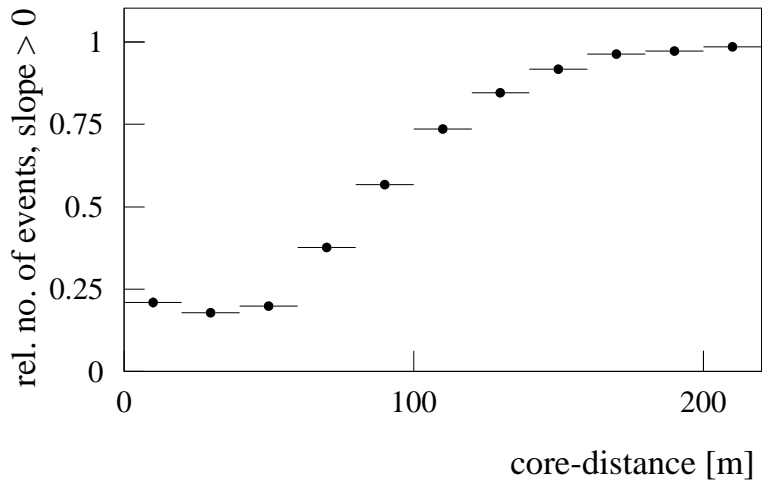}}
\end{center}
\caption
{Fraction of $\gamma$-ray events with a positive time gradient
along $x$, as a function of the core distance.}
\label{fig_tasym}
\end{figure}

The time gradient along the major axis of Cherenkov images can
be used to determine where the head and tail ends of the image are,
at least if the core distance is approximately known. To
study how reliable such a technique is, a linear time gradient 
was fitted to the individual images, taking into account the estimated errors
for the timing values of the individual pixels. 
Fig.~\ref{fig_tasym} illustrates what
fraction of the $\gamma$-ray images have a positive gradient, i.e. d$t$/d$x > 0$,
as a function of the core distance. As already evident from 
Fig.~\ref{fig_long2}, events with small core distances tend to have
negative slopes, events with large distances have positive slopes.
For distances less than about 60~m and for distances greater than 100~m,
the orientation of the image can be determined with 75\% reliability. 
For stereoscopic
systems, this information is of course of limited use, since the stereoscopic
reconstruction will  anyway resolve the usual head-tail ambiguities.

Information concerning the orientation of images is much more relevant
for single Cherenkov telecopes. Based on the results shown in
Fig.~\ref{fig_tasym}, it is however nontrivial to make statements about the
performance in the single telescope case, because there the information
on impact distance is poorer and also the distribution of impact
distances will be different than for the IACT system.
%

Image timing information has been repeatedly proposed for use in 
$\gamma$-hadron discrimination. Hadron images should show a larger timing
spread \cite{time1}, and in particular "precursor" signals at early times caused
by Cherenkov light emitted from high-energy muons \cite{time1,time2}.

There are two different approaches to using the timing information: 
one option is to
use the full signal waveform, as recorded by the Flash-ADC system, 
before or after deconvolution. In this case, one would typically sum over 
the waveform for all image pixels, and then calculate the width of the pulse
or look for precursors or a tail in the waveform. One could also
consider the pulse widths measured for the individual pixels. An alternative is to use
the timing values derived for each pixel, and to calculate the width of the
distribution, or look for early or late pixels. With a sampling rate of the
Flash-ADCs of 120 MHz, one is obviously not very sensitive to differences
in pulse shapes at the level of at most a few ns, as they are expected 
between $\gamma$-ray images and cosmic-ray images. Work therefore concentrated
on the spread of the light arrival times measured in the
 different pixels of an image; here, the
experimental resolution is in the relevant (ns) range.

\begin{figure}[htb]
\begin{center}
\mbox{
\epsfxsize14.0cm
\epsffile{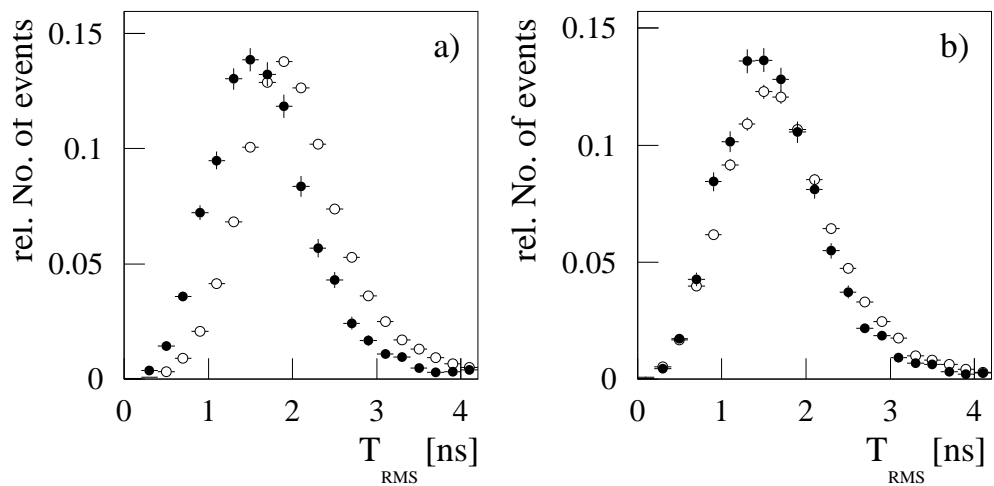}}
\end{center}
\caption
{Distribution of the rms spread per image of pixel times, relative
to the nominal time profile expected for $\gamma$-ray events. Open symbols
indicate cosmic-ray events, closed symbols $\gamma$-ray images.
(a) For all events, (b) after selecting $\gamma$-ray candidates 
on the basis on image shape.}
\label{fig_discr}
\end{figure}

Several different approaches
were tried. For example, Fig.~\ref{fig_discr}(a) shows the rms spread of
pixel times with respect to the time profile expected for $\gamma$-ray images.
One rms value is calculated using all pixels of a given image, and the
distribution of this rms spread is plotted for $\gamma$-ray images and
for cosmic-ray background events. Cosmic-ray events clearly show a 
larger spread among pixel times, compared to $\gamma$-ray images.
However, the differences are not significant enough to allow an 
efficient separation; at most, a Q-factor of around 1.1 is reached.

In addition, the timing information is highly correlated with the
image shapes. Once $\gamma$-ray candidates are selected on the basis
of cuts on the mean scaled {\em width} of all images, 
the time spread among image 
pixels is almost identical for $\gamma$-ray images and for cosmic
rays (Fig.~\ref{fig_discr}(b)). 

Various other techniques to use the timing information for discrimination
were investigated, with similar results; in no case a reasonably
clear-cut difference between $\gamma$-ray images and cosmic rays was
seen. These results differ from the conclusions obtained in 
\cite{time1}, where an improvement of significance with an analysis of
signal shapes was obtained even after shape cuts. However, the 
hardware setups and analysis techniques differ significantly. Also,
the trigger conditions and 
shape cuts used in case of the HEGRA stereoscopic system provide
a much stronger pre-selection than in case of the single telescope of
~\cite{time1}.

\section{Summary}

Using air showers observed with the stereoscopic HEGRA IACT system,
with their precisely reconstructed direction and core location,
the mean arrival time of Cherenkov photons was studied as a function
of the direction of these photons, i.e., of their location within
the shower image, and as a function of the core distance.
The key feature of cosmic-ray induced images is a roughly linear
variation of arrival times along the major axis of an image. 
The gradient varies almost linearly with the core distance, from
values of about -1 ns/$^\circ$ for small distances to more than 
5 ns/$^\circ$ for large distances. In
the transverse direction, along the minor axis, a parabolic profile
is observed, with photons on the major axis of the image arriving earliest.
Gamma-ray images differ from cosmic-ray images in that particles at
small angles -- emitted at large height -- tend to arrive later than
the bulk of the image photons. Only at larger core distances, well
beyond 100~m, does the linear variation along the major axis become
the main feature.

These features of the time profiles of Cherenkov images can be
explained in a semi-quantitative fashion as an interplay between
the speed of propagation of the shower front, and the speed of
propagation of Cherenkov photons on their way from the point of emission
to the telescope.

The observed variation of photon arrival times imposes a lower limit
on the signal integration time used in Cherenkov telescopes; if a common
integration gate is used for all pixels, gate times significantly
below 10~ns will result in a distortion of image parameters, and possibly
in a reduced gamma-hadron separation.

The use of pixel timing to achieve improved $\gamma$-hadron separation and towards
resolving the head-tail ambiguity of the images is discussed. Concerning
$\gamma$-hadron separation, one finds that timing information is
not a very efficient method of hadron rejection. The timing information is strongly
correlated with the information contained in images' shapes; after an 
efficient selection on the basis of shapes, no further improvement was
possible. Timing can be used to resolve head-tail ambiguities to a certain
extent. In the case of the HEGRA IACT system, the timing information
provided by the 120 MHz Flash-ADC system is valuable for consistency
checks and for 
the fine-tuning of the simulation, but does not significantly improve the 
performance of the system. We believe that even with waveform recording at
higher frequencies, the gain is quite limited in the case of IACT systems, since
trigger conditions and shape cuts preselect a `$\gamma$-ray like' sample
of hadronic showers, which also in their timing properties behave much like
$\gamma$-induced showers. The situation may be different for single IACTs,
with their less stringent preselection, see e.g. \cite{time1}.

\section*{Acknowledgements}

The support of the German Ministry for Research 
and Technology BMBF and of the Spanish Research Council
CYCIT is gratefully acknowledged. We thank the Instituto
de Astrofisica de Canarias for the use of the site and
for providing excellent working conditions. We gratefully
acknowledge the technical support staff of Heidelberg,
Kiel, Munich, and Yerevan.

\end{document}

%% file: hegra_appt.tex
\author[1]{M.~He\ss},
\author[1,9]{K.~Bernl\"ohr},
\author[1]{A.~Daum},
\author[1]{M.~Hemberger},
\author[1]{G.~Hermann},
\author[1]{W.~Hofmann},
\author[1]{H.~Lampeitl},
\author[1]{F.A.~Aharonian},
\author[2]{A.G.~Akhperjanian},
\author[3,4]{J.A.~Barrio},
\author[4]{J.J.G.~Beteta},
\author[4]{J.L.~Contreras},
\author[4]{J.~Cortina},
\author[5]{T.~Deckers},
\author[3,4]{J.~Fernandez},
\author[4]{V.~Fonseca},
\author[4]{J.C.~Gonzalez},
\author[7]{G.~Heinzelmann},
\author[1]{A.~Heusler},
\author[6]{H.~Hohl},
\author[2]{I.~Holl},
\author[7]{D.~Horns},
\author[1,2]{R.~Kankanyan},
\author[3]{M.~Kestel},
\author[5]{O.~Kirstein},
\author[1]{C.~K\"ohler},
\author[1]{A.~Konopelko},
\author[3]{H.~Kornmayer},
\author[3]{D.~Kranich},
\author[1]{H.~Krawczynski},
\author[7]{A.~Lindner},
\author[3]{E.~Lorenz},
\author[6]{N.~Magnussen},
\author[6]{H.~Meyer},
\author[3,4,2]{R.~Mirzoyan},
\author[4]{A.~Moralejo},
\author[4]{L.~Padilla},
\author[1]{M.~Panter},
\author[3,6]{D.~Petry},
\author[3]{R.~Plaga},
\author[7]{J.~Prahl},
\author[3]{C.~Prosch},
\author[1]{G.~P\"uhlhofer},
\author[5]{G.~Rauterberg},
\author[6]{W.~Rhode},
\author[7]{A.~R\"ohring},
\author[5]{M.~Samorski},
\author[4]{J.A.~Sanchez},
\author[7]{D.~Schmele},
\author[6]{F.~Schr\"oder},
\author[5]{W.~Stamm},
\author[1]{H.J.~V\"olk},
\author[6]{B.~Wiebel-Sooth},
\author[1]{C.A.~Wiedner},
\author[5]{M.~Willmer}
 
\collab{HEGRA Collaboration}

\address[1]{Max-Planck-Institut f\"ur Kernphysik, P.O. Box 103980,
        D-69029 Heidelberg, Germany}
\address[2]{Yerevan Physics Institute, Yerevan, Armenia}
\address[3]{Max-Planck-Institut f\"ur Physik, F\"ohringer Ring 6,
        D-80805 M\"unchen, Germany}
\address[4]{Facultad de Ciencias Fisicas, Universidad Complutense,
         E-28040 Madrid, Spain}
\address[5]{Universit\"at Kiel, Inst. f\"ur Kernphysik,
       Olshausenstr.40, D-24118 Kiel, Germany}
\address[6]{BUGH Wuppertal, Fachbereich Physik, Gau\ss str.20,
        D-42119 Wuppertal, Germany}
\address[7]{Universit\"at Hamburg, II. Inst. f\"ur Experimentalphysik,
       Luruper Chaussee 149, D-22761 Hamburg, Germany}
\address[9]{Now at Forschungszentrum Karlsruhe, P.O. Box 3640, 76021 Karlsruhe}
\address[8]{Now at Department of Physics University of Leeds, 
       Leeds LJ2 9JT, UK}